\documentclass{moriond}

\def\Journal#1#2#3#4{{#1} {\bf #2}, #3 (#4)}

\def\PRD{{\em Phys. Rev.} D}

\def\MNRAS{\em Mon. Not. Roy. Astron. Soc.}
\def\LRR{\em Living Rev. Rel.}

\def\be{\begin{equation}}
\def\ee{\end{equation}}
\def\bea{\begin{eqnarray}}
\def\eea{\end{eqnarray}}

\newcommand{\Photo}{}

\begin{document}
\vspace*{4cm}
\title{What's in LISA data?\\End-to-end simulation and analysis pipeline for LISA}

\author{Jean-Baptiste Bayle, Christian Chapman-Bird, Graham Woan}
\address{University of Glasgow,\\Glasgow G12 8QQ, United Kingdom}

\author{Olaf Hartwig, Aur\'elien Hees, Marc Lilley, Peter Wolf}
\address{SYRTE, Observatoire de Paris, Universit\'e  PSL,\\CNRS, Sorbonne Universit\'e, LNE,\\61 avenue de l’Observatoire, 75014  Paris, France}

\maketitle\abstracts{
The data produced by the future space-based millihertz gravitational-wave detector LISA will require nontrivial pre-processing, which might affect the science results. It is crucial to demonstrate the feasibility of such processing algorithms and assess their performance and impact on the science. We are building an end-to-end pipeline that includes state-of-the-art simulations and noise reduction algorithms. The simulations must include a detailed model of the full measurement chain, capturing the main features that affect the instrument performance and processing algorithms. In particular, we include in these simulations, for the first time, proper relativistic treatment of reference frames with realistic numerically-optimized orbits; a model for onboard clocks and clock synchronization measurements; proper modeling of total laser frequencies, including laser locking, frequency planning and Doppler shifts; and a better treatment of onboard processing. Using these simulated data, we show that our pipeline is able to reduce the most critical noises and form synchronized observables. By injecting signals from a verification binary, we demonstrate that good parameter estimation can be obtained on this more realistic setup, extending existing results from previous LISA Data Challenges\cite{ldc}.}

\section{Introduction}

The gravitational-wave detector LISA is planned for launch in the 2030s\cite{proposal}. The raw telemetered (L0) data LISA will produce cannot be directly used for scientific analysis. Instead, it will be processed on ground to produce the noise-reduced, calibrated, clock-synchronized L1 data, from which gravitational waves are extracted. To prepare for mission adoption later this year, we are building a demonstration pipeline that includes simulation of L0 data and a first version of the noise-reduction and calibration algorithms. The aim of this work is to demonstrate that key processing steps can be assembled together and work as expected. In addition, we perform simple parameter estimation on L1 data, and check that results are not impacted by the increased complexity of our processing algorithms.

\section{Orbits}

LISA is made up of 3 spacecraft (labelled 1, 2, 3) in an almost-equilateral triangular formation around the Sun, with a armlength of 2.5 million km. The spacecraft positions and inter-spacecraft light travel times define the nonstationary response of the instrument to gravitational-wave signals. As a first study, we use equal-armlength orbits obtained with a first-order expansion of the Kepler equations, optimized to keep the constellation stable\cite{chevineau+05,lisaorbits}. Nevertheless, we have built our pipeline to accommodate any set of realistic orbits, e.g., provided by ESA.

Physics in each spacecraft is simulated according to the proper time of the spacecraft $\tau_i$ (TPS). As a consequence, we need to relate physical quantities when they are exchanged between spacecraft. Therefore, we also compute the deviations between TPSs and the global Barycentric Coordinate time $t$ (TCB).

\section{Gravitational-wave response}

We compute the strain time series $h_+, h_\times$ and the time-domain response $y$ of each LISA link using LISA GW Response\cite{lisagwresponse}. Link responses can be computed for any strain time series. Here, we choose signals derived from verification binaries, compact stars with ultrashort orbital periods, already observed electromagnetically\cite{kupfer+pre} (in particular, V407Vul with a 569-s period and an integrated 4-year SNR of 44 to 80).

We simulate 3 days of data. To perform parameter estimation, we boost the amplitude of our signal such that the integrated SNR matches the full mission 4-year SNR.

\section{Instrument}

LISA Instrument\cite{bayle+23,lisainstrument} is used to simulate the propagation of modulated optical beams (signals and noises), clocks, phasemeter measurements, onboard computers, and ultimately the quantities telemetered to the ground.

\subsection{Time frames and constellation}

Each spacecraft hosts two laser sources, a phasemeter, a clock driving the phasemeter, and two movable optical sub-assemblies (MOSAs) that we label $ij$, where $i$ is the hosting spacecraft index and $j$ the spacecraft the MOSA points to (see Fig.~\ref{fig:optical-setup}). A MOSA contains a telescope that sends and receive light to and from the other spacecraft, an optical bench with three interferometers, a gravitational reference sensor (GRS) and its test mass (acting as the inertial reference).

The clocks onboard the spacecraft define three time frames $\hat\tau_i$ that are used for all onboard measurements. They differ from the TPSs $\tau_i$ by the instrumental imperfections of the clocks.

\subsection{Optical setup and simulation}

\begin{figure}
    \centering
    \includegraphics[width=\linewidth]{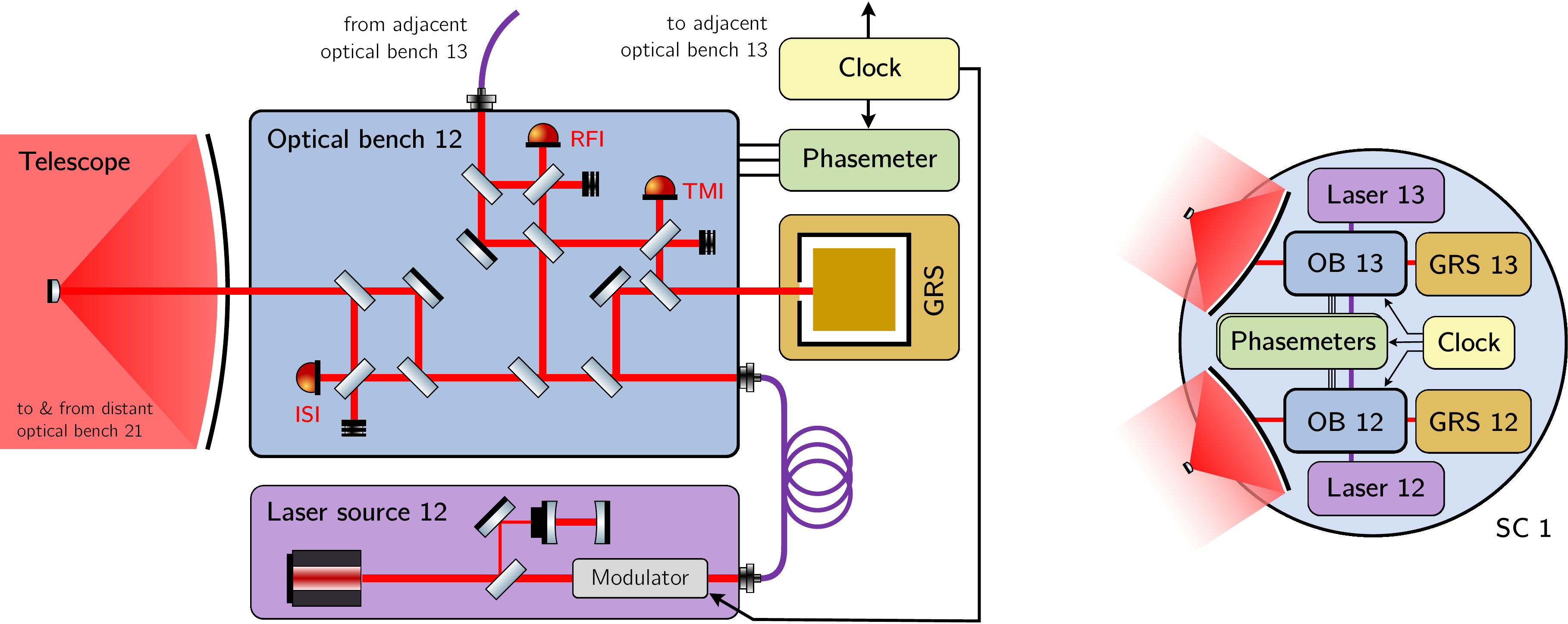}
    \caption{On the right, overview of the some subsystems on a spacecraft. On the left, a closer look at the subsystems associated with one optical bench.}
    \label{fig:optical-setup}
\end{figure}

Since the actual spacecraft and payload design is still under consideration, we use a simplified model, depicted in Fig.~\ref{fig:optical-setup}. This model includes a laser source that can be stabilized on a cavity; the produced beam is described by its electromagnetic field $E(\tau) = E_0(\tau) \cos(2\pi \Phi(t))$, of which we only simulate the instantaneous frequency $\nu(\tau) = \dot\Phi(\nu) / 2\pi$. We can decompose this total frequency into
\begin{equation}
    \nu(\tau) = \nu_0 + \nu^o(\tau) + \nu^\epsilon(\tau),
\end{equation}
where $\nu_0$ $\approx$ 280 THz is the constant central frequency, and $\nu^o$ models MHz Doppler shifts and programmed offsets. Finally, $\nu^\epsilon$ contains instrumental noise, with a magnitude up to $\approx$ 100 Hz, and the small gravitational wave signals at $\approx$ 100 nHz.

The laser beams are injected on the optical bench and guided to three interferometers, the adjacent optical bench, and the telescope. In addition, incoming beams from the telescope and the adjacent optical bench are also guided to the same three interferometers: the inter-spacecraft interferometer (ISI) mixes the distant and local lasers; the reference interferometer (RFI) compares the two adjacent lasers; the test-mass interferometer (TMI) is similar to the RFI, but additionally bounces the local beam on the test mass. The interfering beams at each of these interferometers create a beatnote oscillating at the difference frequency of the two beams, $\nu_1 - \nu_2 \approx$ 10 MHz. These interferometers are connected to the phasemeter to produce the main measurements. Note that gravitational waves are encoded as hundreds of nHz fluctuations in these MHz beatnotes.

The main expected noise source is the inherent frequency instability of the lasers. In addition, the laser beams pick up noises while they are propagated to the interferometers, such as noises due to external forces on the test masses, imperfections of the fiber link between adjacent optical benches, or thermal expansion of the optical bench. The beatnote is then read out by the phasemeter and associated electronics, which also adds additional noise (including the fundamentally limiting shot noise).

\subsection{Onboard processing}

\begin{figure}
    \centering
    \includegraphics[width=\linewidth]{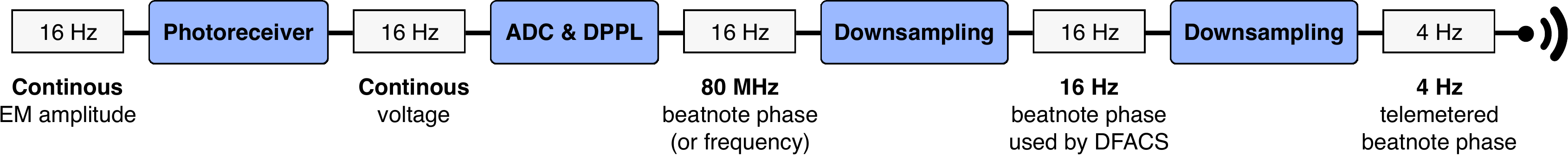}
    \caption{Various sampling rates used in our simulations, compared to reality.}
    \label{fig:sampling-rates}
\end{figure}

All quantities described up to this point are, in reality, continuous. The phase readout, on the other hand, is performed using a digital phasemeter with multiple internal sampling rates ranging from 80 MHz to 4 Hz associated with multiple filtering stages. In our simulations, we use only two sampling rates: 16 Hz for continuous analog and high-frequency digital signals, and 4 Hz for the final telemetry data (see Fig.~\ref{fig:sampling-rates}). We include a single downsampling and filtering step to avoid aliasing.

The phasemeter bandwidth is limited to between 5 and 25 MHz. Because of the time-varying MHz Doppler shifts, we must actively control the offsets between the laser frequencies to ensure all beatnotes fall within that range at all times. This is enforced by locking the lasers onto each other according to a frequency plan, with one primary laser locked to its cavity. The problem of finding such frequency plans is non-trivial, but has recently been solved using methods based on computational geometry\cite{heinzel18}. The resulting pre-computed frequency plans are used as input to the simulation. 

\begin{figure}
    \centering
    \includegraphics[width=\linewidth]{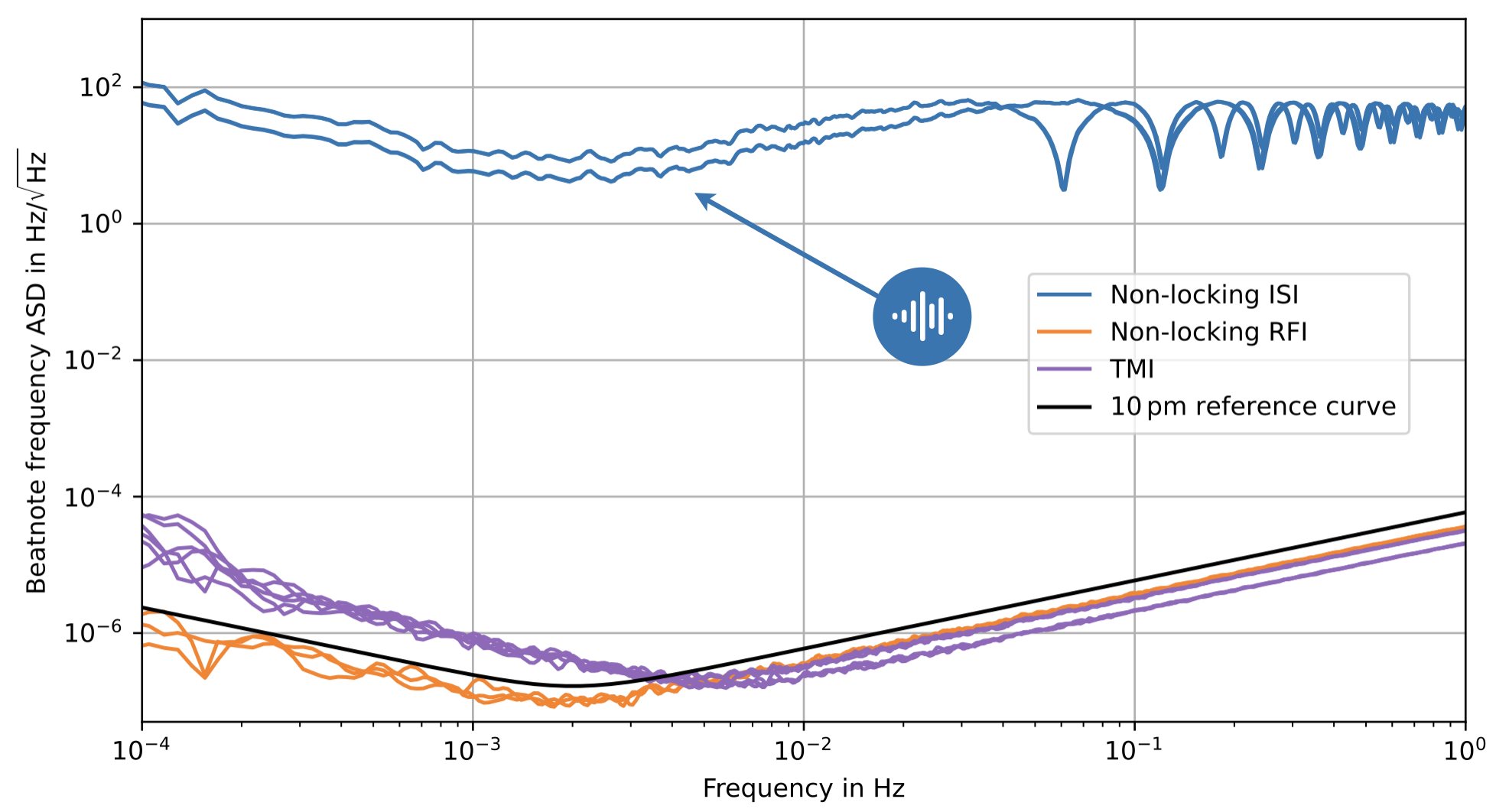}
    \caption{Amplitude spectral densities of the non-locking ISI and RFI beatnotes, as well as all TMI beatnotes. Locking beatnotes are at numerical precision and not shown here.}
    \label{fig:L0-spectra}
\end{figure}

A consequence of laser locking is that any noise in the locking beatnotes will be imprinted on the locked lasers. Therefore, assuming perfect locking, some beatnotes will only contain the slow-varying MHz offsets while non-locking beatnotes in addition contain all noises and gravitational signals. Spectra of these beatnotes are shown in Fig.~\ref{fig:L0-spectra}.

\subsection{Instrumental clock effects}

Detecting 100 nHz signals in MHz beatnotes would require 40 $\mathrm{fs}\,\mathrm{Hz}^{-0.5}$ timing precision clocks. Current space-qualified oscillators fail to meet this requirement by several orders of magnitude. Instead, LISA will use dedicated differential clock measurements to correct clock noise on-ground.

The laser beams are phase modulated to imprint the in-band part of the clock jitter,
\begin{equation}
    E(\tau) = E_0 e^{j 2\pi (\Phi_c(\tau) + m \Phi_m(\tau))}.
\end{equation}
Here, $\Phi_c$ is the total carrier phase discussed above, $m$ is the modulation index, and $\Phi_m$ is the clock-derived modulating signal. The modulated signal can be expanded using Bessel functions into the sum of a carrier and two sideband beams $\approx$ 2.4 GHz away from the carrier. In each interferometer, the sidebands create additional beatnotes, which are tracked by the phasemeter and used on-ground to correct for clock noise.

In addition to this GHz modulation, the carrier is also modulated with a pseudo-random noise (PRN) code. Correlating the PRN codes present in the different beams, which are generated according to different clocks, we can measure the absolute pseudoranges between the spacecraft. The pseudoranges contain the light travel times and the differential clock noises. Combining sideband, PRN and ground-based observations allows to reconstruct a precise and accurate estimate of the pseudorange.

\subsection{Simulation products}

The main simulation products used as input for further processing are the interferometric carrier and sideband beatnote frequencies as well as the PRN measurements.

\section{Ground tracking}

Earth-based observation of the LISA satellite allows to measure their positions and velocities, which is used for on-ground processing (see below). In addition, it is possible to determine the relation between the onboard clock times and the global TCB by comparing the data packets’ onboard timestamps and their UTC reception times. As a simplification, we suppose here that all these quantities are recovered without any errors. This assumption will be relaxed in future studies.

This ground tracking information in conjunction with the main simulation products constitutes our simulated L0 data.

\section{On-ground processing}
\label{sec:onground-processing}

The telemetered phasemeter beatnotes contain the gravitational-wave signals, but also several overwhelming noises. In addition, they are given as functions of the different onboard clock times, as those are not actively synchronized in flight. Therefore, the measurements must be processed on ground to reduce these noises and re-synchronize all data to a global time frame before they can be used for gravitational-wave detection and analysis.

Multiple approaches to on-ground processing have been proposed. We have implemented recent algorithms for laser and clock noise reduction that operate directly on the raw total beatnote frequencies\cite{bayle+21,hartwig+22}. The final processing will contain additional noise-reduction steps.

\begin{itemize}
    \item \textbf{Ranging sensor fusion} – Combine sideband and PRN measurements to estimate the delays (and their derivatives) appearing in the TDI equations.
    \item \textbf{Estimation of light travel times} – Use ground-tracking observations to estimate inter-spacecraft light travel times, required to compute the instrument response function during parameter estimation.
    \item \textbf{TDI XYZ} – Compute second-generation time-delay interferometry (TDI) Michelson combinations\cite{tinto+21} to suppress the overwhelming laser and clock noises\cite{hartwig+22}.
    \item \textbf{Re-synchronization} – Use clock information from ground-tracking to re-synchronize TDI XYZ to a global time frame (TCB).
    \item \textbf{TDI AET} – Compute quasi-orthogonal TDI combinations used for faster likelihood evaluation during parameter estimation.
\end{itemize}

\begin{figure}
    \centering
    \includegraphics[width=\linewidth]{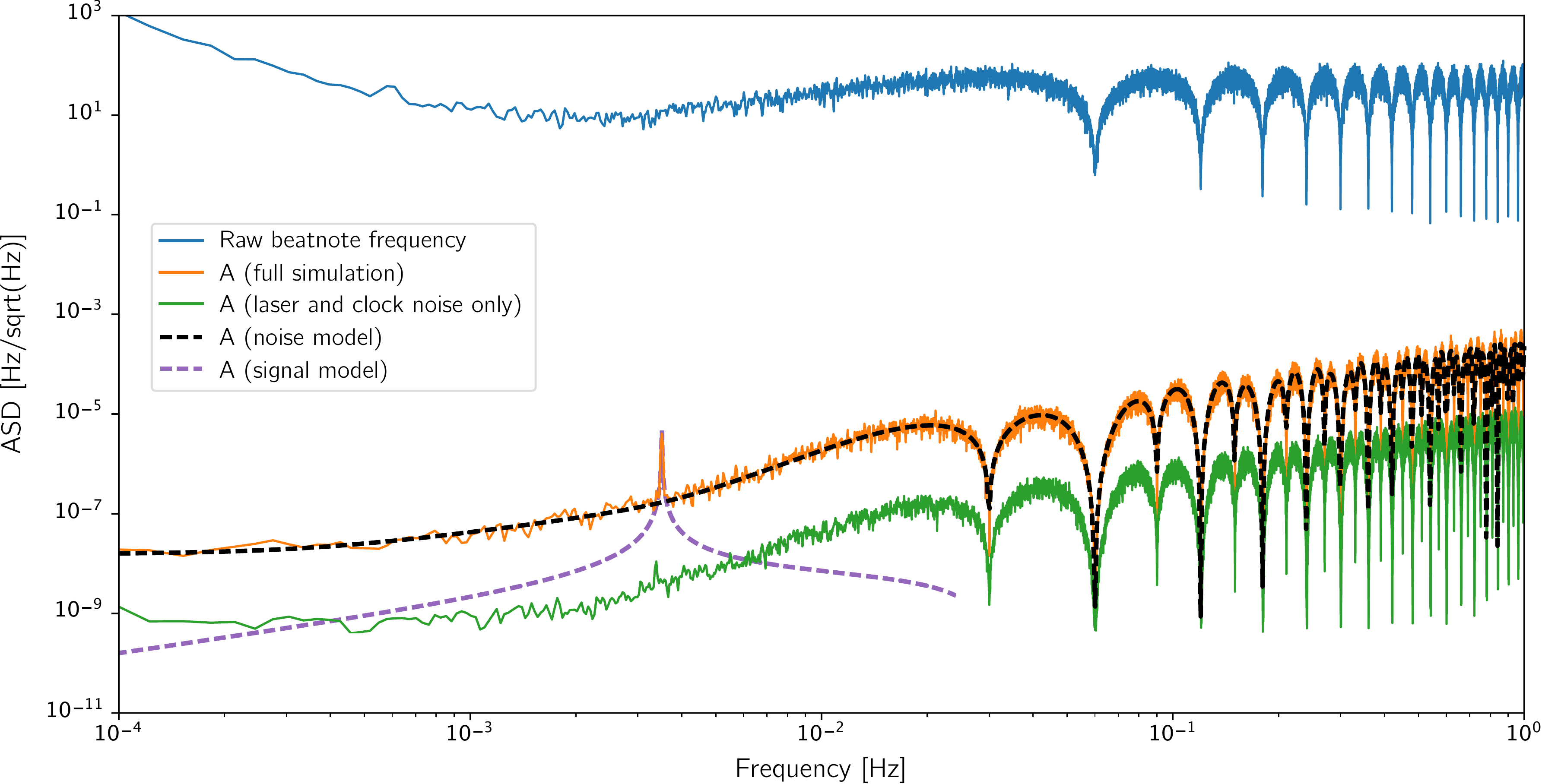}
    \caption{Spectrum of the raw ISI beatnote frequency, dominated by laser noise (in blue). Pre-processing quasi-orthogonal $AET$ TDI combinations (in orange) have laser noise reduced by more than 8 orders of magnitude (in green), and are now dominated by secondary noises (noise model as the black dashed line). The injected GW signal is now visible (signal model as the purple dashed line).}
    \label{fig:spectrum}
\end{figure}

Fig.~\ref{fig:spectrum} shows the resulting L1 data spectrum. The injected signal clearly appears in $A$, which is now dominated by readout and test-mass acceleration noises after laser and clock noises have been suppressed.

\section{Parameter estimation}

Parameter estimation is performed on the frequency-domain re-synchronized TDI combinations, using a Gaussian likelihood based on noise (the noise covariance matrix only contains the dominant test-mass and shot noises) and signal (using the frequency-domain \texttt{FastGB} template generation method\cite{cornish07}).

\begin{figure}
\begin{minipage}{0.5\linewidth}
\centerline{\includegraphics[width=\linewidth]{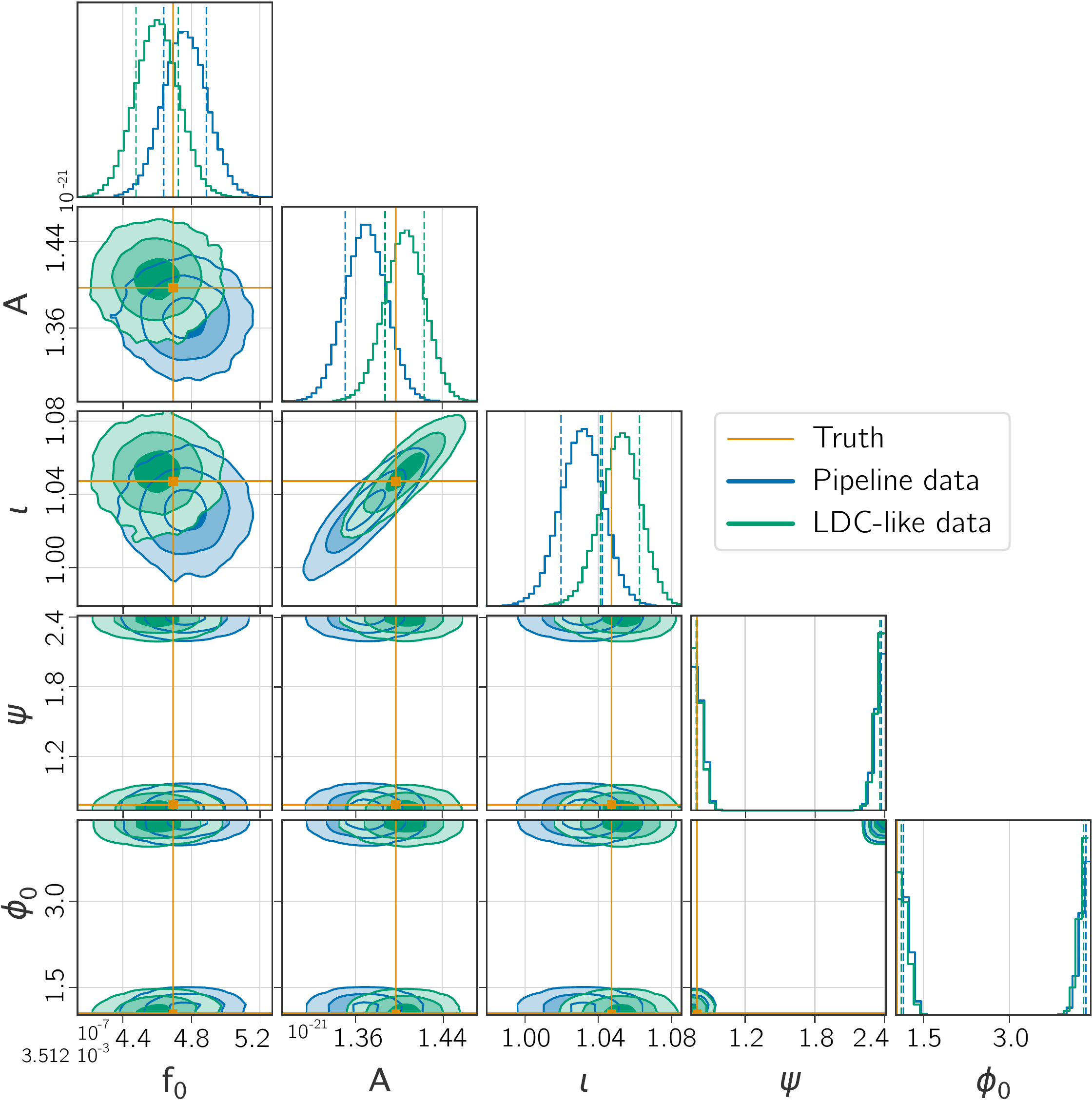}}
\end{minipage}
\hfill
\begin{minipage}{0.5\linewidth}
\centerline{\includegraphics[width=0.9\linewidth]{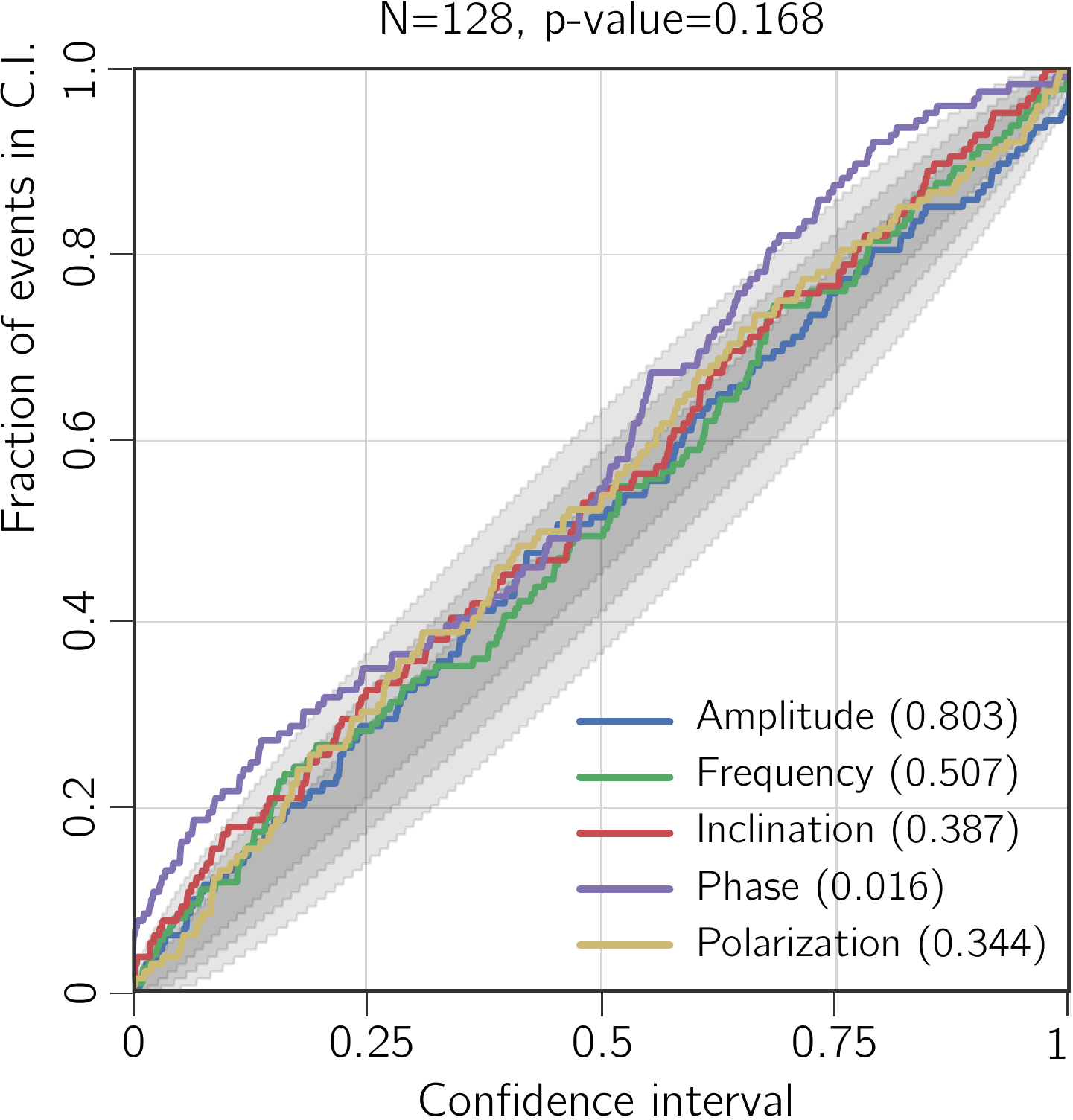}}
\end{minipage}
\hfill
\caption{On the left, posterior distributions obtained with Nessai for the injected gravitational-wave signal (frequency and amplitude $f_0$ and $A$, inclination $\iota$, polarization $\psi$, and initial phase $\phi_0$) for our demonstration pipeline (in blue) and a simpler LDC-like pipeline (in green). On the right, p-p plot for 128 realizations using our pipeline.}
\label{fig:param-estimation}
\end{figure}

We used two samplers (Nessai\cite{williams+21} and Dynesty\cite{speagle20}), and recovered all parameters correctly with the exception of the frequency derivative and sky localization angles, which cannot be recovered due to the short simulation length. Comparing posteriors obtained with the data generated by our pipeline against posteriors from simpler LDC-like data\cite{ldc}, which ignores most of the complexity described above, we see no significant degradation in parameter estimation. The corner plot on the left of Fig.~\ref{fig:param-estimation} compares the results of a likelihood analysis performed using Nessai on the data from our pipeline with those of an analysis performed on the LDC-like data. Note that it does so for two different noise realizations. The good agreement of the posteriors shows that on-ground pre-processing successfully suppresses and corrects for instrumental artifacts, and that a likelihood analysis model that does not account for these remains valid.

The p-p plot shown on the right of Fig.~\ref{fig:param-estimation} below further confirms that this simple likelihood model is mostly consistent with the full simulated data. We see a slight deviation for the initial phase, which is fully explained by additional delays due to the filtering performed in the simulation we did not account for in this study.

We plan to extend this work by correcting for filter delays, and use several month-long simulations to include sky localization parameters. In addition, we plan to upgrade the processing pipeline to include further noise reduction steps, such as tilt-to-length coupling corrections.

\section*{Acknowledgments}

J.B.B. gratefully acknowledges support from UK Space Agency (grant ST/X002136/1). O.H., A.H., M. L. and P.W. gratefully acknowledge support by Centre National d'\'Etudes Spatiales (CNES) for the LISA mission. O.H. acknowledges support by the Deutsches Zentrum für Luft- und Raumfahrt (DLR, German Space Agency) with funding from the Federal Ministry for Economic Affairs and Energy based on a resolution of the German Bundestag (Project Ref.~No.~50OQ1601 and 50OQ1801). This work was supported by the Programme National GRAM of CNRS/INSU with INP and IN2P3 co-funded by CNES.

\end{document}